\documentclass[runningheads]{svmult}

\usepackage{makeidx}   
\usepackage{graphicx}  
\usepackage{subeqnar}  
\usepackage{multicol}  
\usepackage{physmubb}  
\makeindex             

\usepackage{psfig}
\usepackage{epsfig}
\usepackage{amssymb,wasysym,bbm,feynmf,rotating,subfigure,here,citesort}

\newcommand{\s}[1]{\mbox{Sec\_$\:\!$#1}}   
\newcommand{\xyz}{\linebreak[4]}
\newcommand{\be}{\begin{equation}}
\newcommand{\ee}{\end{equation}}
\newcommand{\ww}[1]{\mbox{$\,#1\,$}}
\newcommand{\www}[1]{\mbox{$\;#1\;$}}
\newcommand{\bea}{\begin{eqnarray}}
\newcommand{\eea}{\end{eqnarray}}
\newcommand{\benn}{\begin{displaymath}}
\newcommand{\eenn}{\end{displaymath}}
\newcommand{\beann}{\begin{eqnarray*}}
\newcommand{\eeann}{\end{eqnarray*}}
\newcommand{\barray}{\begin{array}}
\newcommand{\earray}{\end{array}}
\newcommand{\w}{\frac}

\newcommand{\x}{\large}

\newcommand{\yyy}{\scriptsize}

\newcommand{\z}{\mbox}
\newcommand{\zz}{\stackrel}
\newcommand{\zzz}{\hspace}

\newcommand{\m}{\z{$\;\!\!$}}       
\newcommand{\mm}{\z{$\:\!\!$}}      
\newcommand{\mmm}{\z{$\!$}}         
\newcommand{\0}{\z{$\,\!$}}         
\newcommand{\1}{\z{$\:\!$}}         
\newcommand{\2}{\z{$\;\!$}}         
\newcommand{\3}{\z{$\,$}}           
\newcommand{\4}{\3\1}               
\newcommand{\5}{\3\2}               
\newcommand{\7}{\4\3}               
\newcommand{\8}{\5\3}               
\newcommand{\di}[1]{${\renewcommand{\0}{\renewcommand}
\0{\2}{\z{2$\:\!$-dimensional}}
\0{\3}{\z{3$\:\!$-dimensional}}
\0{\4}{\z{4$\;\!$-dimensional}}
\0{\7}{\z{7-dimensional}}
\mbox{#1}}$}                      
\newcommand{\DI}[1]{\z{$D=#1$}}   
\newcommand{\DIM}[1]{\z{($\,D=#1\,$)}}  
\newcommand{\four}{4$\;\!$-}      
\newcommand{\gtsim}{\;\lower-0.45ex\hbox{$>$}\kern-0.77em\lower0.55ex\hbox{$\sim$}\;}
\newcommand{\ltsim}{\;\lower-0.45ex\hbox{$<$}\kern-0.77em\lower0.55ex\hbox{$\sim$}\;}
\newcommand{\trace}{\z{trace}} 
\newcommand{\diff}{\z{d}}   
\newcommand{\diffA}{\z{d}\1\mathcal{A}} 
\newcommand{\vecA}{\1\vec{\mathcal{A}}} 
\newcommand{\diffV}{\z{d}\mathcal{V}}   
\newcommand{\q}[1]{\z{\x\ttfamily\itshape\m#1\4}} 
\newcommand{\A}{\q{\1A\mmm}} 
\newcommand{\F}{\q{\1F\mm}}  
\newcommand{\g}{\z{\m\ttfamily\itshape g\1}}
\newcommand{\tp}{\z{${\2}^{\z{\yyy t}}\0$}}
\newcommand{\befig}{\begin{figure}}
\newcommand{\efig}{\end{figure}}
\newcommand{\betab}{\begin{table}}
\newcommand{\etab}{\end{table}}
\begin{document}
%
%
%

\begin{titlepage}
\begin{flushright}
\begin{tabular}{l}
HD-THEP-03-10\\
\end{tabular}
\end{flushright}

\vspace*{1.3truecm}

\begin{center}
\boldmath
{\Large \bf Self-Similarity between 3-dimensional Magnetostatics 
and 4-dimensional Electrodynamics}\\ 
\vspace*{0.2truecm}
\unboldmath

\vspace*{1.6cm}

\smallskip
\begin{center}
{\sc 
{\large J.~Holk$^\diamond$
}}\\
\vspace*{2mm}
{\sl Institut f\"ur Theoretische Physik, Universit\"at Heidelberg\\
Philosophenweg 16, D-69120 Heidelberg, Germany}
\end{center}

\vspace{2.0truecm}

{\large\bf Abstract\\[10pt]} \parbox[t]{\textwidth}{ 
  It is shown that Euclidean electrodynamics is the exact  
  4-dimensional ana- logue of 3-dimensional magnetostatics. 
  This concept is related to a 4-dimen- sional generalization 
  of the cross product between two vectors where the only 
  essential modification is made to the tensor rank of the 
  involved arguments. Parallels to the Schl\"afli-Coxeter 
  theory of Platonic polytopes are pointed out.
}

\end{center}

\vskip 0.5 in

\footnoterule
\vskip 3truemm
{\small\tt
\noindent$^\diamond$J.Holk@thphys.uni-heidelberg.de}

\end{titlepage}
 
\thispagestyle{empty}
\vbox{}
\newpage
 
\setcounter{page}{1}
 
%
%
%
%
%
%
%
\titlerunning{Self-Similarity between 3-dimensional Magnetostatics 
and 4-dimensional Electrodynamics}
%
%
%


\section{Introduction}
\label{Sec_Introduction}

In a recent paper by Silagadze~\cite{Silagadze:2001ri}, an alternative 
derivation of the Maxwell equations is given which mixes classical and 
quantum concepts. Furthermore, that author presents a \di{\7}  
generalization of the cross product between two vectors where the 
structure constants are by part admitted to be zero, even in case 
of all three indices being different from each other. 

It is well-kown that velocity and acceleration are the first and the 
second derivative of the co\"ordinate vector relative to time and that 
these relations are remaining true at the level of the Euclidean 
representation of special relativity if the co\"ordinate vector is 
replaced by the event vector and time is replaced by proper time. In 
this paper, we will show that exactly this receipt also transforms 
magnetostatics into electrodynamics, without further assumptions to be 
performed. This procedure is associated with a \di{\4} (formula  
language) translation of the \di{\3} cross product between two vectors 
which makes a compromise on the tensor rank of the linked variables, 
but not on the Levi-Civita character of its structure constants.

This paper is organized as follows. In \s{2}, we are preparing the 
mathematical foundations for the reinterpretation of special relativity 
in the context of self-similarity, to be analysed in \s{3}. The  
problematic around the energy-momentum tensor is discussed in \s{4}. The 
final section is devoted to interdisciplinary perspectives.

\section{Exterior Derivatives and the Cross Product}
\label{Sec_Algebraic_Definitions}

Let \www{\vec{b}=\vec{b}(\vec{x})} denote an arbitrary \di{\3} real 
co\"ordinate-dependent vector field. The tensorial translation (in the 
sense of formula language) of its exterior derivative 
\be
        \diff(\vec{b}\cdot\diff\vec{x})
        =
        \z{curl}\,
        \vec{b}\cdot\z{d}\vecA    
\ee
--with 
\www{\z{curl}\,\vec{b}=\varepsilon^{ijk}\,\hat{\vec{e}}_{\1i}\,
\partial_j\,b_k}  
(using sum convention in three dimensions, the Euclidean Levi-Civita tensor 
\www{\varepsilon_{ijk}=\varepsilon^{ijk}}, and canonical base vectors 
\ww{\hat{\vec{e}}_{\1i}} with 
\www{\hat{\vec{e}}_{\1i}\mm\cdot\m\vec{b}
= b_{\1i}\8\forall\4i}) 
and the vectored areal element
\www{\diff\vecA=\varepsilon^{mnr}\,\hat{\vec{e}}_m\,
\diff x_n\wedge\diff x_r} 
(turning to the normal of the differential of area--and with calligraphic 
$\vecA$~,\xyz in order to distinguish it from alphabetically similar 
figurative sym-\xyz bols which will be worked with below, like \z{"\q{A}"} 
and \z{"$\!A\,$"\1)}--is known as\xyz  
\www{
        \diff(b_i\,\diff x^i)
        =
        \w{1}{2}\,
        (\z{CURL}\,b)_{jk}\,
        \diff x^{\1j}\wedge\diff x^k
}, 
with the components of the tensorial CURL (we are using the ad-hoc 
notation "CURL" rather than "Curl" in order to facilitate the optical 
distinction between "curl" and "Curl") given by  
\www{(\z{CURL}\,b)_{ij}=\partial_{\1i}\,b_j-\partial_j\,b_{\1i}}. 
Indeed the CURL is not only feasible in three dimensions, but can also be 
comfortably generalized to four dimensions (cf. def. (26) below), belonging 
to the most genuine second differential form there, as the curl is in three 
dimensions. 

Therefore the general understanding up to now is that the \di{\4} CURL is 
the \di{\4} analogue of the \di{\3} curl. The intention of this paper is to 
demonstrate that this interpretation guides into an impasse in that way 
that it conceals the aspects presented below.

The counter-suggestion is to assign the label "curl" to the third differential 
form in four dimensions. Let \ww{C_{\alpha\beta}} with 
\www{C_{\alpha\beta}=-\,C_{\beta\alpha}} 
be a totally antisymmetric tensor of rank 2 in real \z{\four space.} Then 
it has the same tensor transformation properties (just the given 
antisymmetry) as any result of a CURL in real \z{\four space.} The third 
differential form in four dimensions can now be written as 
\be
        \diff(C_{\alpha\beta}\,\diff x^\alpha\wedge\diff x^{\1\beta})
        =
        (\z{curl}\,C)_\gamma
        \,
        \diffA^{\m\zz{\z{\yyy$\1\gamma$}}{}}  
\ee
with 
\www{\z{curl}\,C:=\varepsilon^{\alpha\beta\gamma\delta}\,
\hat{e}_\alpha\,\partial_\beta\,C_{\gamma\delta}} 
and the element of \di{\3} hypersurface "area" in \z{\four space}
\www{\diffA:=\varepsilon^{\mu\nu\m\rho\sigma}\,\hat{e}_\mu\,
\diff x_\nu\wedge\diff x_\rho\wedge\diff x_\sigma},  
where again\xyz   
\www{\varepsilon_{\alpha\beta\gamma\delta}=
\varepsilon^{\alpha\beta\gamma\delta}} 
is the Euclidean representation of the Levi-Civita tensor in four 
dimensions (it is rather trivial to hint at the existence of a 
corresponding integral version according to the general integral 
theorem of 
Stokes 
\www{       
        \int_M
        \diff\omega
        =
        \oint_{\2\partial\m M}
        \omega
},
where \ww{\omega} is any \z{(\ww{p}-1)} form being continuously differentiable 
in an open superset of any compact orientated smooth \z{$p\2$-dimensional} 
submanifold $M$ in \z{$D\1$-dimensional} flat space, with boundary 
\ww{\2\partial\m M} in induced orientation--just identify 
\www{
        \omega
        =
        C_{\alpha\beta}\,\diff x^\alpha\wedge\diff x^{\1\beta}  
}). 

If we demand that 
\be
        \z{curl}\,C
        =
        \partial
        \times
        C    
\ee
in four dimensions, analogically to 
\be
        \z{curl}\,\vec{b}
        =
        \vec{\partial}
        \times\1
        \vec{b}    
\ee
in three dimensions, then the \di{\4} cross product acts on a \z{4\2-dimen-} 
sional vector $b$ and an antisymmetric \di{\4} dyad $C$ according to 
\be
        b\times C
        :=
        \varepsilon^{\alpha\beta\gamma\delta}
        \,
        \hat{e}_\alpha
        \,
        b_\beta
        \,
        C_{\gamma\delta}
        \ ,    
\ee
also adhering to the \di{\4} Euclidean version of the Levi-Civita tensor 
\www{\varepsilon_{\alpha\beta\gamma\delta}=
\varepsilon^{\alpha\beta\gamma\delta}}.  
This definition is self-consistent in every respect and will be motivated 
physically in the next section (including the geometric concept 
of orthogonality, which remains applicable in \z{\four space}--it will be 
enlighted in the discussion of the \di{\4} reinterpretation of 
chain of eq. (18)~). For this purpose, we finally need further auxiliary 
definitions for the ensuing differential operators, acting on vectors $b$ 
and antisymmetric dyads $C$ in real \z{\four space:} 
\www{
        \diff(b_\alpha\,\diffA^{
        \zz{\z{\yyy$\alpha$}}{}})
        =
        \z{div}\,b\8
        \diffV
} 
with 
\be
        \z{div}\,b  
        :=
        \partial_\beta
        \,
        b^{\2\beta} 
\ee
and Euclidean \di{\4} 
\www{\diffV=
\diff x_{\zz{}{1}}\wedge
\diff x_{\zz{}{2}}\wedge
\diff x_{\zz{}{3}}\wedge
\diff x_{\mm\zz{}{4}}},
\be
        \z{div}\,C  
        :=
        \partial_\alpha
        \,
        C^{\1\alpha\beta} 
        \,
        \hat{e}_\beta
        \ ,
\ee
as well as (the following definition is motivated by the insertion 
of a unit dyad w.r.t. (3) and (5), supplying one index for the 
argument of the \di{\4} cross product and the other index as supplement 
to the result of that cross product, where $1/2$ is a compensating 
normalization factor)
\be
        (\z{curl}\,b)_{\alpha\beta}
        :=
        \w{1}{2}
        \,  
        \varepsilon_{\alpha\beta\gamma\delta}
        \,
        \partial^{\m\zz{\z{\yyy$\1\gamma$}}{}}
        \,
        b^{\1\delta}
\ee
so that the \di{\4} curl converts an antisymmetric dyad into a vector, 
and vice versa. Please realize that the \di{\4} Levi-Civita tensor 
behaves anti-cyclic relative to an exchange of its indices, unlike its 
\di{\3} counterpart. All these relations can and will be taken over for 
the Euclidean representation of special relativistic spacetime, where 
both the metrics and its inverse are equivalent to a Kronecker Delta 
tensor, but the fourth component of any physical vector exhibits to be 
purely imaginary.

\section{A Novel Access to Special Relativity}
\label{Sec_Special_Relativity}

In order to facilitate the comparison between \di{\3} space and \z{4\2-di-}  
mensional Euclidean spacetime, we will reserve ourselves to drop the tensor 
specification--or the vector bold type, respectively--when regarding 
\z{3\1-dimen-} sional objects, from now on as already done in case 
of \z{\four space} objects, where a corresponding convention exists 
anyway. This means that we want by intention to leave any tensor of any 
rank ambiguous with respect to its relevant dimensionality (~\DI{3} for 
space and \DI{4} for Euclidean spacetime) if and only if its contingent 
indices are omitted. This concept will always leave undecided whether a 
scalar has to be treated as an invariant of classical mechanics (or other 
nonrelativistic physics, else) or as an invariant of special relativity, 
which can indeed differ by a velocity-dependent factor. 

We start (here as well as later on, the anticipated superscript $\1\tp\3$ 
designates a transposition, which turns a row vector into a column vector,  
e.g.) with the co\"ordinate  vector, that then changes from 
\be
        x
        =
        \tp(
        x_{\zz{}{1}},
        x_{\zz{}{2}}\1,
        x_{\zz{}{3}}
        \1)
\ee
to 
\be
        x
        =
        \tp(
        x_{\zz{}{1}},
        x_{\zz{}{2}}\1,
        x_{\zz{}{3}}\1,
        x_{\mm\zz{}{4}}
        \1)\ ,
        \zzz{.5cm}
        x_{\mm\zz{}{4}}
        =
        i\,c\:t
\ee
if space \DIM{3} is replaced by Euclidean spacetime \DIM{4}. In both cases, we
want to define a generalized length \z{$L(x)$} by the main value of the square 
root 
\be
        L(x)
        :=
        \sqrt{
        \tp
        x
        \cdot
        x
        \;
        }\ |
        ^{\z{\yyy main value}}
        \ , 
\ee
which is selected by fixing the complex phase angle \ww{\z{arg}(L(x))} within 
the interval \ww{]\!-\1\!\pi,\pi\1]}. It is obvious that \z{$L(x)$} is the real 
geometric length in the event of \DI{3} (referring to space), and something 
like a pseudo-length in the event of \DI{4} (referring to Euclidean spacetime: 
the imaginary unit \ww{i} in (10) shall be taken seriously in the framework of 
this topic), which is nevertheless conserved strictly w.r.t. all 
transformations of the system of reference there. In both cases (~\DI{3\,} and 
\z{$\,$\DI{4}~),} the associated (actual or generalized, respectively) scalar 
product is bilinear (in formal excess of sesquilinearity for \z{\DI{4}~!\2)} 
and equivalent to the corresponding clear-cut matrix product: 
\www{
        x
        \mm\cdot\m 
        y
        =
        \tp\m
        x
        \m\cdot
        y 
        =
        \tp
        y
        \m\cdot
        x
}. 
Note that \z{$L(x)$} can become purely imaginary for 
\DI{4}~.

Now we are able to specify time in both scenarios 
(~\DI{3\,} and \z{$\,$\DI{4}~).} The time $T$ alters from normal time 
\be
        T
        =
        \2
        t 
\ee
for \DI{3} to proper time 
\be
        T
        =
        -\,i
        \int
        L
        (\diff x)
        \;
        \z{sgn}
        (\diff\1t\1)
        \2/\,
        c
\ee
for \DI{4}~, with \z{$L(\diff x)$} relating to \DI{4} of course (but 
\z{d$\1t\1$} still refers to normal time \ww{t}). 

The formulae (9) up to (13) are supporting the specifications of velocity 
\be
        V
        =
        \w
        {\diff\1x}
        {\diff T}
\ee
and acceleration 
\be
        \A
        =
        \w
        {\diff^2x}
        {\diff T^2}
\ee
(italic non-bold typewriter face for distinction from \z{"$\!A\,$",} which  
will be defined later) simultaneously for both scenarios of reference 
(~\m\DI{3\,} and \z{$\,$\DI{4}~).} We designate this behaviour as 
non-Mandelbrot self-similarity and we will use the underlying principle 
to rediscover special relativity by the reinterpretation of the most 
fundamental laws of classical mechanics and magnetostatics, employing (9) 
up to (13). 

In a first tentative step, we try to extrapolate the definition of 
momentum 
\be
        P
        =
        m
        \,
        \w
        {\diff\1x}
        {\diff T}
\ee
and the (simplest mechanical one-body description for the) equations of 
motion for force  
\be
        \F
        =
        m
        \,
        \w
        {\diff^2x}
        {\diff T^2}
\ee
(italic non-bold typewriter face for distinction from \z{"\mm$F_{\mu\nu}$\1",} 
which will be defined later) to special relativity, taking advantage of the 
pre-results in (14) and (15). If we interpret \ww{m} as normal mass for \DI{3} 
and as the relativistically (i.e.: w.r.t. special relativity) invariant rest 
mass for \DI{4} and if we demand that $P$ and $\F$ have to be 
\z{\four momentum} and \z{\four force} in Euclidean special relativity, then 
in fact (16) and (17) remain true for \DI{4}~!  

In electrostatics, the electric field is constant relative to time (in the 
scope of the probe time interval of reference) and no (macroscopically 
relevant) magnetic field does appear. We formally define magnetostatics by 
the complementary situation that the magnetic field is static and that there 
is effectively no non-zero electric field present in a given system. Imagine 
the situation inside a Faraday cage surrounding circuits of constant direct 
current (coils are admitted--we are thinking of the idealized classical 
picture of a co-moving probe charge $q$ with mass $m$~), or discard the 
entity of the Lorentz force (skipping (18) and (19), v.i.) and just 
refer to the most superficial description of magnetism between permanent 
magnets, for example. 

Regarding this magnetostatic type of situation, we can argue that 
\be
        \w
        {\diff}
        {\diff T}
        \,
        L^{\12}
        (V)
        =
        \w
        {\diff}
        {\diff T}
        \,
        V^{\12}
        =
        2\cdot\!V\!\cdot\mm\A\1
        =
        \w{2}{m}\,
        V\!\cdot\mm\F\1
        =
        2\,
        \w{q}{m}\,
        V\!\cdot(V{\times}B)
        =
        0
        \ ,
\ee
where (11), (14), (15), and (17) are applied for \DI{3}~, specifying $\F$                                                         to be the magnetostatic (static relative to the magnetic field \z{$B$~,} 
not relative to \z{\ww{x\,}:} $V$ does not vanish in general) Lorentz 
force (in syst\`{e}me international d'unit\'{e}s, SI, or with 
\z{$c=\m1$~)}
\be
        \F
        =
        q\,
        V{\times}B
        \ .
\ee
(18) and (19) (within the scope of \z{(18)\3)} are telling us that 
\be
        L(V)
        =
        \z{const}
\ee
so that $V$ can only change its direction, which can be handled as a process 
of pure revolving of $V$ itself, with \ww{\w{\zz{\z{\yyy$q$}}{}}{m}B} being 
the angular velocity pseudotensor of the associated rotation of \z{$V$~.} The 
rank of that angular velocity pseudotensor \ww{\w{\zz{\z{\yyy$q$}}{}}{m}B} 
corresponds to the number of simultaneously possible linearly independent 
momentary planes of rotation. It is clear that this rank is one for \DI{3}~, 
confirming that the magnetic field is a pseudovector. On the other hand, 
the \DI{4} analogue for $B$ has to be an angular velocity pseudotensor of the 
second rank for the same reason, being realized by an antisymmetric dyad. 

From (11) and (14), we can deduce that 
\www{L(V)=i\,c} 
for \DI{4}~, with \ww{i} being the imaginary unit and \ww{c} being the 
velocity of light (which will be set equal to unity subsequently) because 
(10) and (13) are supplying contributions to (14) which are not totally 
independent of each other. This implies that (20) is always fulfilled for  
\DI{4}~, referring to full electrodynamics of special relativity (indeed, 
this aspect is relevant for the whole domain of special relativity, 
including relativistic mechanics as well--for the ensuing illustration, we 
will, however, focus upon the picture of electrodynamics)! 

We would hence expect that there is an angular velocity representation 
for acceleration or (the comparison is rendered possible by \z{(17)\3)}
force like (19) for \DI{4}~, concerning the \di{\4} interpretation of the 
cross product given in (5). This is in fact possible if we choose $\F$ 
to be the \z{\four vector} formulation of the complete electrodynamic 
Lorentz force, \ww{q} to be the relativistically (i.e.: w.r.t. special 
relativity) invariant rest charge, $V$ to be given by (14), and $B$ to be 
the antisymmetric dyad with the components 
\be
        B_{\m\mu\nu}
        =
        \w{1}{2}
        \,
        \tilde{F}_{\mu\nu}
        =
        \w{1}{4}
        \,
        \varepsilon_{\mu\nu\m\rho\sigma}
        \,
        F^{\rho\sigma}
        \ ,
\ee
where \z{$F_{\mu\nu}$} and \z{$\tilde{F}_{\mu\nu}$} are the actual and the dual 
field strength tensor in Euclidean special relativity, respectively. Because of 
(5), all steps taken in (18) are valid for \DI{4} as well, telling us that $V$ 
is perpendicular to \z{($\,V{\times}B\:$)}, independent of the dimensionality  
of reference (~\DI{3\,} or \z{$\,$\DI{4}~).}

In order to check the self-consistency of the preceding considerations, we 
can investigate what the \di{\4} (formula language) translation of the 
magnetostatic differential relations 
\be
        \z{div}\,
        B
        =
        0
\ee
and (the field equations for a field representation relative to the 
classical vacuum, using CGS units with \z{$c=\m1$} and that redefinition 
of the current density which gets rid of the insignificant conversion 
factor \z{$(4\2\pi)$~)}
\be
        \z{curl}\,
        B
        =
        j
\ee
are resulting in if we take \ww{j} to be the special relativistic 
\z{\four current} density (in corresponding normalization) and if 
we evaluate the auxiliary relationships (2), (3), (5), (7), (10), and 
(21). This means that we continue applying the principle of 
self-similarity. The resulting \DI{4} interpretations of the equations 
(22) and (23) can be retranslated into relationships between the common 
physical quantities (tensors of rank 0 and 1 since objects like a stress 
tensor are not relevant here) of the nonrelativistic 
description of electrodynamics, referring to time and \di{\3} position 
space. In this manner, we 
obatain the full homogeneous Maxwell equations from (22) and the full 
inhomogeneous Maxwell equations from \z{(23)\2!} According to the fact that 
the repeated application of an exterior derivative is zero in flat space, 
we can automatically derive the continuity equation 
\be
        \z{div}\,
        j
        =
        \z{div}\;
        \z{curl}\,
        B
        \equiv
        0
        \ ,
\ee
both for nonrelativistic magnetostatics \DIM{3} and for relativistic 
electrodynamics \DIM{4}. 

The integral versions 
\www{       
        \oint
        B
        \mm\cdot\m
        \diffA
        =
        \10
} 
and 
\www{       
        \oint
        B
        \mm\cdot\m
        \diff s
        =
        \int
        j
        \mm\cdot\m
        \diffA
} 
of (22) and (23) are valid for \DI{3} and \DI{4} if we apply the general version 
of Stokes' theorem as mentioned in \s{2} (using the \z{\four definition} of 
\ww{\diffA} given there, 
\www{
        B
        \mm\cdot\m
        \diff s 
        :=
        B_{\alpha\beta}
        \,
        \diff x^\alpha\wedge\diff x^{\1\beta} 
}, 
and 
\www{
        B
        \mm\cdot\m
        \diffA 
        :=
        B_{\alpha\beta}
        \;
        \hat{e}^{\1\alpha}
        \,
        \diffA^{\1\beta}
} 
for 
\DI{4}~). The adaption of the general integral theorem of Stokes to Euclidean 
special relativity with (10) will not provide anything useful, but is exempt from 
problems. Don't confound it with its notorious special formulation for \di{\2} 
submanifolds of \di{\3} position space in nonrelativistic physics, to be appplied 
for (23) in the event of \DI{3}~.

The lemma of Poincar\'{e} predicts that (22) is implying a potential 
representation
\be
        B
        =
        \mbox{curl}\,
        A
\ee
for \DI{3} and sufficiently mathematically benign \z{$B$~.} Although we can 
immediately convince ourselves that (25) remains valid if \ww{A} is reinterpreted 
as the standard \z{\four potential} in Euclidean special relativity and $B$ 
and $\;$curl$\;$ are viewed in the context of (21) and (8), the way of arguing 
has to be modified since the genuine  $\;$curl$\;$ in four dimensions is given 
by (2), and not by (8). We can improve the situation by explicitly introducing 
the second differential form 
\be
        \diff(b_\alpha\,\diff x^\alpha)
        =
        (\z{chi}\,b)_{\mu\nu}\,
        \diff x^\mu\wedge\diff x^\nu
\ee
with 
\www{(\z{chi}\,b)_{\mu\nu}
:=
\w{1}{2}\,(\partial_{\m\mu}\,b_\nu -\partial_\nu\,b_\mu)}
in \z{\four space} (assigning the less misapprehensive expression "chiasm" 
to half the tensorial CURL in four dimensions) and the \di{\4} extension  
\be
        E_{\mu\nu}
        =
        \w{i}{2}
        \,
        F_{\mu\nu}
\ee
of the electric field \z{$E_{\1m}$~,} where \ww{i} is the imaginary unit and 
\z{$F_{\mu\nu}$} is again the field strength tensor in Euclidean special 
relativity. Then (2), (21), (22), and (27) are yielding 
\www{
        \z{curl}\,
        E
        =
        0
}, 
which is implying 
\www{
        E
        =
        \z{chi}\,
        (i\2A)
} 
according to the lemma of Poincar\'{e} (again, \ww{i} is the imaginary 
unit) and can be converted into (25), using (8), (21), (26), and (27). We 
can conclude that (22) is implying (25) for \DI{4} if both the lemma of 
Poincar\'{e} and the duality of $E$ and $B$ are utilized. For \DI{3} on 
the other hand, $E$ and $B$ are not dual relative to each other, and thus 
merely the lemma of Poincar\'{e} is required for deriving (25). 

It is remarkable that the demonstrated analogy between magnetostatics in 
three dimensions and Euclidean electrodynamics in four dimensions does fit 
up to the tiniest detail. Hence this non-Mandelbrot type of self-similarity 
concept is a powerful and quick tool for checking the normalization of 
conversion factors in Minkowski spacetime since a remodeling of metrics 
is in most cases less sophisticated than infinitesimal calculations in 
relativistic mechanics, as well as in electrodynamics.

\section{Visiting General Relativity}
\label{Sec_General_Relativity}

The r\^{o}le of the symmetric energy-momentum tensor in electrodynamics 
is far less clear. Let us denote it by \z{$\Theta_{\mu\nu}^{(2)}$}~, for 
reasons explained below. If we use the \di{\4} Euclidean 
electric and magnetic fields defined in (21) and (27) we can write
\be
        \Theta_{\mu\nu}^{(2)}
        =
        -\,
        (
        E_{\mu\rho}
        \,
        E_{\sigma\nu}
        +
        B_{\m\mu\rho}
        \,
        B_{\m\sigma\nu}
        )
        \,
        g^{\rho\sigma}
        \sqrt{g\,}
        \ ,
\ee
where \ww{g} is the determinant of the metric tensor 
\z{$g_{\zz{}{\mu\nu}}$~.} \ww{\sqrt{g\,}} is used for 
sakes of completeness (in the framework of a density-like definition 
for any energy-momentum tensor, like here) only since it is equal to 
unity in Euclidean special relativistic electrodynamics.

Electrodynamics is based on the unitary group U(1), whose Lie algebra has 
one generator only, given by the scalar factor \z{$2^{-0.5}$} if the same 
normalization standard is applied as in any unitary group or subgroup (i.e. 
any actual or special orthogonal or unitary group). By generalizing (28) 
to any Yang-Mills theory, we obtain an equation which can be formally
expressed in terms of 
\be
        \Theta_{\mu\nu}^{(k)}
        =
        \left(
        \delta_{\!\mu}^\alpha
        \,
        g^{\beta\gamma}
        \,
        \delta_{\m\nu}^{\1\delta}
        +
        g_{\!\zz{}{\zz{}{\mu\nu}}}
        \,
        \w{
        g^{\alpha\gamma}
        \,
        g^{\beta\delta}
        }{2\,k}
        \right)
        \!
        \sqrt{g\,}
        \;
        \Omega_{\alpha\beta\gamma\delta}^{(k)}
\ee
with 
\be
        \Omega_{\alpha\beta\gamma\delta}^{(k)}
        =
        -\,
        \Omega_{\beta\alpha\gamma\delta}^{(k)}
        =
        -\,
        \Omega_{\alpha\beta\delta\gamma}^{(k)}
        =
        \Omega_{\gamma\delta\alpha\beta}^{(k)}
        \zzz{.6cm}
        \forall
        \zzz{.3cm}
        \alpha,\beta,\gamma,\delta 
        \ ,
\ee
using \z{$k=2$} and 
\be
        \Omega_{\alpha\beta\gamma\delta}^{(2)}
        =
        \w{1}{2}
        \,
        \trace
        \,
        (
        F_{\alpha\beta}
        \,
        F_{\gamma\delta}
        )
        \ .
\ee

The formulae (29) and (30) are structurally reminiscent of the Einstein 
field equations in general relativity. These are given by (29) for 
\z{$k=1$} if 
\be
        \Omega_{\alpha\beta\gamma\delta}^{(1)}
        =
        \w{1}{\kappa}
        \,
        R_{\alpha\beta\gamma\delta}
        \ ,
\ee
using the Einstein gravitational constant \z{$\kappa$~,} the Riemann 
tensor \z{$R_{\alpha\beta\gamma\delta}$~,} and identifying 
\z{$\Theta_{\mu\nu}^{(1)}$} with the standard energy-momentum 
tensor of general relativity in \z{$(i\,c\:t)\1$-Euclidean} 
formulation (spacetime is curved, but the fourth component of any 
physical rank one tensor is chosen to be purely imaginary). Furthermore, 
(30) is fulfilled for \z{$k=1$} w.r.t. (32), too.

It is possible to introduce an SO(4) field strength tensor in classical 
\z{$(i\,c\:t)\1$-Euclidean} general relativity via 
\be
        F_{\mu\nu}
        =
        \w{1}{i\,\g}
        \,
        [
        \2
        D_{\mm\mu}
        \2,
        D_\nu
        \2
        ]_{\2\zz{}{\z{-}}}
        \zzz{.6cm}
        \z{with}
        \zzz{.3cm}
        D_{\mm\mu}
        =
        \partial_\mu
        +
        i\,\g\,
        A_\mu
        \ ,
\ee
like in U(1) or SU(N) if the true physical coupling \ww{\g} (this is 
not the same \z{"\mm$g$\3"} as in (28) or (29)--the charge \ww{q} of 
an electron is equal to \ww{\g} in case of U(1) electrodynamics, 
corresponding to the square root of the fine structure constant 
there) is replaced by the artificially fixed pseudo-coupling
\be
        \g\1
        \equiv
        \z{const}
        =
        -\,2
\ee
and 
\be
        A_\mu
        =
        \w{i}{4}
        \:
        {\q{V}}^{a\1\alpha}
        \;
        {\q{V}}^b_{\;\;\alpha\1;\1\mu}
        \left(
        \1
        \hat{e}_a
        \!\cdot\!
        \tp\hat{e}_{\zz{}{b}}
        -
        \hat{e}_{\zz{}{b}}
        \!\cdot\!
        \tp\hat{e}_a
        \right)
        \ ,
\ee
where \ww{i} is the imaginary unit, \z{$\hat{e}_a$} are the basis 
vectors of the locally associated Frenetian 
vierbein~\cite{D'Adda:1991ua,Duan:1996in,Nakamichi:1991ym,Regge:ed.kt}, 
\z{${\q{V}}^a_{\;\;\alpha}$} (having no reference to \z{"$V$\,",} as 
it is given by \z{(14)\3)} is an indexed representation of the vierbein 
transformation matrices w.r.t. Cartan base vectors 
\z{$\hat{e}_\alpha$~,} and the semicolon marks the covariant derivative 
referring to the following adjacent index in the sense of classical 
general relativity. By sticking to (33) up to (35), we can extend the 
definition (31) to be relevant for general relativity, with (30) 
remaining true for \z{$k=2$} in this situation. Specially (no pun, 
allusion, or reference to "special relaitivity") for general relativity 
then 
\be
        \Omega_{\alpha\beta\gamma\delta}^{(2)}
        =
        \w{\kappa^2}{2}
        \:
        \Omega_{\alpha\beta\mu\nu}^{(1)}
        \;
        g^{\mu\rho}
        \,
        g^{\nu\sigma}
        \,
        \Omega_{\rho\sigma\gamma\delta}^{(1)}
        \ ,
\ee
demonstrating a quasi-quadratic relationship.

(29), (30), and (36) are illuminating the closest structural coherencies 
between the self-similarity scheme of \s{3} and general relativity at the 
level of classical (macroscopic) physics. One should be aware of the 
circumstance that (29) is specifying the physical ingredients of the 
regarded energy-momentum tensor for \z{$k=2$~,} while it is rendering 
equations of motion when \z{$k=1$~.}

\section{Concluding Remarks}
\label{Sec_Conclusion}

The discussion of non-Mandelbrot self-similarity in \s{3} is 
engendering the impression that several of the most fundamental 
laws of classical mechanics and magnetostatics, regarding 
our normal \di{\3} space, are repeated once more at the higher 
level of Euclidean special relativity if the product of the 
imaginary unit, the velocity of light, and our normal time is added 
as a fourth dimension. Therefore the laws of nonrelativistic 
mechanics and magnetostatics can be considered as \di{\3} copies 
of perfectly analogous hyper-laws in \di{\4} space. Apparently, 
nature is allocating resembling r\^{o}les to \di{\3} and \di{\4} 
space.

It is striking to diagnose a similar phenomenon in aesthetics. The 
\z{$D\1$-di-} mensional generalization of a polyhedron is called a 
polytope. A \z{$D\1$-dimen-} sional Platonic (introducing the 
required generalization for this conception just hereby) polytope 
is homogeneously bounded by facets in the same style and size, being 
\z{$(D\!-\!1)\1$-dimensional} Platonic polytopes, if \z{$D\ge1$~.} We 
have to distinguish between regular Platonic polytopes which are 
convex and related constructions, being classified as Platonic star 
polytopes (by defining a regular star polytope to be irregular in 
the sense of a normal convex polytope, for reasons of clarity). For 
instance, both pentagon and pentagram are homogeneously surrounded 
by five lines of the same size each, with the line being the unique 
Platonic polytope in one dimension. It is intuitively evident that 
the pentagon is a \di{\2} regular Platonic polytope and that the 
pentagram is a \di{\2} Platonic star polytope. 

The entirety of all Platonic polytopes has been predicted theoretically 
by Schl\"afli in the 19th century and visualized up to dimensionality 
five by Coxeter in the 20th century~\cite{Conway:math}. Let 
\z{$N_{\:\!\!s}(D)$} be the number of all  different Platonic polytopes 
in $D$ dimensions including star polytopes, \z{$N_{\;\!\!p}(D)$} be 
the corresponding number for regular
 Platonic polytopes exclusively, and let $D$ be a natural number 
in this context. According to the Schl\"afli-Coxeter theory, 
we obtain \www{N_{\:\!\!s}(D)=D^{\12}} and \www{N_{\;\!\!p}(D)=D+2} 
both for \DI{3} and for \DI{4}~, while 
\www{N_{\:\!\!s}(D)=N_{\;\!\!p}(D)} for all other dimensionalities 
$D$~, with \www{N_{\;\!\!p}(0)=N_{\;\!\!p}(1)=1}, 
\www{N_{\;\!\!p}(2)=+\,\infty}, and \www{N_{\;\!\!p}(D)=3} 
for all \z{$D>4$}~. Again, we observe a distinct similarity between 
\di{\3} and \di{\4} space.

%

%

\end{document}